\documentclass[10pt,journal,compsoc]{IEEEtran}
\usepackage{amsmath}    
\usepackage{amssymb}    
\usepackage{amsfonts}   
\usepackage{graphicx}
\usepackage{subfig}

\begin{document}
\flushbottom  

\title{Dynamic-Attention-based EEG State Transition Modeling for Emotion Recognition}
\author{Xinke Shen\textsuperscript{†}, Runmin Gan\textsuperscript{†}, Kaixuan Wang, Shuyi Yang, Qingzhu Zhang, Quanying Liu, Dan Zhang*,~\IEEEmembership{Member,~IEEE,} Sen Song*%
\thanks{\textsuperscript{†}These authors contributed equally to this work.}%
\thanks{Xinke Shen, Qingzhu Zhang and Quanying Liu are with the Department of Biomedical Engineering, Southern University of Science and Technology, Shenzhen 518055, China.}%
\thanks{Runmin Gan, Shuyi Yang and Sen Song are with the School of Biomedical Engineering, Tsinghua University, Beijing 100084, China and Tsinghua Laboratory of Brain and Intelligence, Tsinghua University, Beijing 100084, China.}%
\thanks{Kaixuan Wang is with the School of Artificial Intelligence, Beijing Normal University, Beijing, 100875, China and Tsinghua Laboratory of Brain and Intelligence, Tsinghua University, Beijing 100084, China.}%
\thanks{Dan Zhang is with the Department of Psychological and Cognitive Sciences, Tsinghua University, Beijing 100084, China and Tsinghua Laboratory of Brain and Intelligence, Tsinghua University, Beijing 100084, China.}%
\thanks{Sen Song and Dan Zhang are the corresponding authors.}%
}

\markboth{Journal of \LaTeX\ Class Files,~Vol.~14, No.~8, August~2015}%
{Shen \MakeLowercase{\textit{et al.}}: Dynamic-Attention-based EEG State Transition Modeling for Emotion Recognition}

\IEEEtitleabstractindextext{%
\begin{abstract}
Electroencephalogram (EEG)-based emotion decoding can objectively quantify people's emotional state and has broad application prospects in human-computer interaction and early detection of emotional disorders. Recently emerging deep learning architectures have significantly improved the performance of EEG emotion decoding. However, existing methods still fall short of fully capturing the complex spatiotemporal dynamics of neural signals, which are crucial for representing emotion processing. This study proposes a Dynamic-Attention-based EEG State Transition (DAEST) modeling method to characterize EEG spatiotemporal dynamics. The model extracts spatiotemporal components of EEG that represent multiple parallel neural processes and estimates dynamic attention weights on these components to capture transitions in brain states. The model is optimized within a contrastive learning framework for cross-subject emotion recognition. The proposed method achieved state-of-the-art performance on three publicly available datasets: FACED, SEED, and SEED-V. It achieved $75.4\pm5.5\%$ accuracy in the binary classification of positive and negative emotions and $59.3\pm7.7\%$ in nine-class discrete emotion classification on the FACED dataset, $88.1\pm3.6\%$ in the three-class classification of positive, negative, and neutral emotions on the SEED dataset, and $73.6\pm12.7\%$ in five-class discrete emotion classification on the SEED-V dataset. The learned EEG spatiotemporal patterns and dynamic transition properties offer valuable insights into neural dynamics underlying emotion processing.
\end{abstract}

}

\maketitle

\IEEEdisplaynontitleabstractindextext

\IEEEpeerreviewmaketitle

\IEEEraisesectionheading{\section{Introduction}\label{sec:introduction}}

%
%
%
%
\IEEEPARstart{E}{motions}  play a crucial role in human social life. Quantifying human emotions and enabling intelligent devices to interpret them can significantly enhance human-centric intelligent interaction systems. This also facilitates real-time monitoring and early detection of emotional disorders\cite{picard2000affective}. Electroencephalography (EEG) measures multi-channel electrical activities of the neural system, providing rich information and high objectivity of emotion recognition, which has increasingly attracted researchers' attention\cite{shu2018review}.

Many studies have explored effective EEG features and deep learning architectures for emotion decoding. Commonly used features include power spectral density (PSD)\cite{koelstra2011deap} and differential entropy (DE, equivalent to the logarithmic energy spectrum)\cite{duan2013differential}, both of which capture the power or energy of local electrode signals in specific frequency bands. Various deep learning methods, such as recurrent neural networks (RNNs)\cite{zhang2018spatial}, convolutional neural networks (CNNs)\cite{Cui2020RegionalAsymmetricCNN}, and graph neural networks (GNNs)\cite{song2018eeg}, have been applied to learn emotion-related representations from DE features. Attention mechanisms have also been adopted recently to emphasize critical feature dimensions for emotion recognition\cite{Tao2020ChannelWiseAttention}. Many of these methods focus on identifying interaction between features from different brain regions. However, the rapid dynamics of EEG signals on finer time scales have been largely ignored.

Neural activities exhibit complex and intriguing spatiotemporal dynamics~\cite{northoff2020temporo}. From the neuron population to the whole-brain level, the spatial and temporal dynamic analyses are key to understanding the brain's organization and functions~\cite{xu2023interacting},~\cite{zhou2013spatiotemporal}. EEG signals encompass orchestration dynamics of neuronal activities at the macro- and millisecond-level~\cite{cohen2017does}. State transition modeling is a major approach to studying EEG spatiotemporal dynamics, which reveals meta-stable states of spatial, spectral, or network activities in EEG~\cite{michel2018eeg},~\cite{vidaurre2016spectrally},~\cite{duc2019microstate}. Temporal dynamics of these transient states serve as functional or behavioral signatures and changes in neurological or mental disorders~\cite{gui2020assessing},~\cite{milz2016functional},~\cite{zappasodi2019eeg},~\cite{guan2022eeg},~\cite{bai2021spontaneous}. Different emotion states modulate temporal features and transition properties of these EEG states~\cite{shen2020exploring},~\cite{chen2021dual},~\cite{prete2022exploring},~\cite{hu2023eeg},~\cite{liu2023eeg}. It is promising to investigate EEG spatiotemporal dynamics to understand how the brain produces different emotion states and develop biophysically interpretable models for emotion recognition. Previous EEG state transition modeling mainly relies on simple clustering methods like microstate~\cite{michel2018eeg} or statistical methods like hidden Markov models~\cite{vidaurre2016spectrally}. These methods suffer from weak representative power and do not account for the inter-subject differences of EEG dynamics. Deep learning provides flexible model architectural design and self-supervised learning strategies to extract target-related and generalizable representations~\cite{liu2021self}. However, current deep-learning-based emotion recognition methods largely overlook EEG state transition properties. A deep learning architecture designed for EEG state transition modeling can potentially boost emotion recognition performance.

In this paper, we propose a Dynamic-Attention-based EEG State Transition (DAEST) model for emotion recognition. With the assumption that dominant EEG spatiotemporal components vary in different brain states, we estimate dynamic attention weights on these spatiotemporal components at each time point. The spatiotemporal components are extracted using a combination of temporal convolution for spectral decomposition and spatial transition convolution for spatial variation pattern estimation. A depthwise convolution with temporal pooling is employed to estimate the dynamic attention weights on these components. We utilize a contrastive learning method for inter-subject alignment in cross-subject emotion recognition. It identifies interpretable emotion-related EEG spatiotemporal dynamics shared across subjects and facilitates the generalization of emotion recognition across different individuals. The code is made publicly available at https://github.com/RunminGan1218/DAEST.

\section{Related work}
\subsection{Modeling of EEG Spatiotemporal Dynamics}

The spatiotemporal dynamics of EEG capture large-scale activities, reflecting fundamental operating patterns of the human brain. Variations in these EEG dynamic features are indicative of different mental states and disorders~\cite{michel2018eeg},~\cite{gui2020assessing},~\cite{milz2016functional}. Distinct emotion states also exhibit characteristic EEG spatiotemporal dynamic features~\cite{schiller2024eeg},~\cite{Hsu2022UnsupervisedLearning}. Several spatiotemporal dynamic modeling methods have been employed in emotion recognition studies.

Microstate analysis identifies several typical EEG activation patterns through spatial clustering. Each pattern remains stable for tens of milliseconds before transitioning to another state\cite{michel2018eeg}. The temporal characteristics of EEG microstates reflect the underlying processes of various emotional states. Microstate C and D are associated with arousal and valence processing in music video watching, respectively\cite{hu2023eeg}. Microstates C and B are most relevant in distinguishing discrete emotions\cite{liu2023eeg}. For emotion regression, microstates in finer frequency bands predict emotions better than those in broadband\cite{shen2020exploring}. Beyond simple spatial pattern clustering, Hsu et al. proposed the adaptive mixture independent component analysis (AMICA) model, assuming different combinations of independent components under various states\cite{Hsu2018AdaptiveMixtureICA}. The EEG states identified by the model in an unsupervised manner correspond to different imagined emotional states\cite{Hsu2022UnsupervisedLearning}. Dynamic brain network methods have also been applied to emotion recognition by modeling the transitions of brain network activations. Yahya et al. enhanced emotion recognition accuracy and sensitivity by integrating local cortical activation and dynamic functional network connectivity\cite{al2019emotion}. These studies underscore the significance of EEG's spatiotemporal dynamics in emotional processing. However, most methods still rely on clustering or statistical learning to model brain state transitions. Leveraging the powerful representation capabilities of neural networks to extract EEG state transition properties can potentially identify more effective emotion representations and improve decoding performance.

Besides, EEG spatiotemporal dynamics have been shown to vary significantly across individuals and are associated with factors such as personality traits and intelligence scores\cite{zanesco2020within}\cite{liu2020reliability}. This variability poses a challenge in cross-subject emotion recognition, making it crucial to address and mitigate these differences in spatiotemporal dynamic features. While statistical methods have been proposed to align hidden states across subjects in fMRI studies\cite{lee2023hyper}\cite{taghia2018uncovering}, there remains a lack of effective approaches for inter-subject spatiotemporal dynamics alignment for EEG data.

\subsection{Deep Learning Architectures for EEG Emotion Recognition}

To enhance the performance of emotion decoding, various deep learning architectures have been explored for EEG-based emotion recognition to capture spatial and/or temporal information from EEG data.

One prominent approach in modeling the spatial relationships across EEG channels is the use of Graph Neural Networks (GNNs). Song et al. treated each EEG channel as a node in a graph and applied graph convolution to features extracted from each channel, effectively integrating information across channels\cite{song2018eeg}. Building on this, subsequent studies introduced sparse constraints to GNNs to better reflect the sparsity of inter-channel relationships\cite{Zhang2021SparseDGCNN},\cite{Zhong2020EEGEmotionRecognition}. Recently, more graph network variants were proposed to further improve the emotion recognition performance\cite{Du2022MultiDimensionalGCN},\cite{Sun2022DualBranchGraph},\cite{Zhang2019GCBNet},\cite{ye2024semi}. For instance, Ye et al. proposed a model that combines GNN and fully connected networks to extract both spatially structured and unstructured information from differential entropy features and employed self-attention mechanisms for multi-
branch feature fusion\cite{ye2024semi}.

Recurrent Neural Networks (RNNs) \cite{zhang2018spatial},\cite{LiY2022HierarchicalSpatialTemporal},\cite{Du2022EfficientLSTM} and Convolutional Neural Networks (CNNs) \cite{Alhagry2017EmotionRecognitionLSTM},\cite{Xing2019SAELSTM},\cite{Cui2020RegionalAsymmetricCNN},\cite{Siddharth2022DeepLearningMultimodal},\cite{Cho2020SpatioTemporalRepresentation} are widely used for extracting both spatial and temporal information from EEG data. Zhang et al. proposed a two-layer RNN to handle both spatial and temporal sequences within EEG\cite{zhang2018spatial}. They transformed EEG channel features into spatial sequences and modeled their dynamics within a time window. Li et al. further expanded on this by introducing hierarchical spatial-temporal RNNs to process local relationships and global relationships progressively\cite{LiY2022HierarchicalSpatialTemporal}. Considering the asymmetry in brain hemisphere activities related to emotions\cite{Hinrikus2009SpectralAsymmetryIndex}, Li et al. suggested using four parallel RNNs to extract spatial sequence information from both brain hemispheres in different directions\cite{LiY2021BiHemisphericDiscrepancy}, performing inter-hemispheric operations such as subtraction and division to highlight inter-hemispheric differences. For CNNs, several studies interpolated EEG electrode distributions into two-dimensional maps, using convolution operations to extract spatial and temporal patterns from EEG features\cite{Cui2020RegionalAsymmetricCNN},\cite{Siddharth2022DeepLearningMultimodal},\cite{Cho2020SpatioTemporalRepresentation}. Cui et al. suggested using two-dimensional convolution to extract local spatial features of EEG and an asymmetry difference layer to capture asymmetry information between the left and right hemispheres\cite{Cui2020RegionalAsymmetricCNN}.

Recently, attention mechanisms have been introduced to further enhance the modeling of crucial spatial and temporal features in EEG data. Tao et al. applied channel attention mechanisms to identify important EEG electrodes and self-attention mechanisms to identify crucial time segments\cite{Tao2020ChannelWiseAttention}. Li et al. expanded this approach by combining deep convolution and Transformer architectures to extract local and global temporal features hierarchically\cite{Li2023MACTN}. Liu et al. introduced a spatiotemporal Transformer framework (EeT), which uses simultaneous spatiotemporal attention mechanisms to learn EEG signal representations, significantly improving emotion recognition accuracy\cite{Liu2023SpatialTemporalTransformers}.

Despite these advances, current deep learning methods still have limitations in capturing the full complexity of EEG dynamics. Most models focus on static spatial or temporal features and do not fully explore the dynamic transitions between different brain states. There is a need for a deep learning architecture that can effectively model the spatiotemporal dynamics of EEG signals to further improve the performance of emotion recognition.

\subsection{Cross-subject EEG Emotion Recognition Methods}
Domain adaptation methods have been widely adopted to tackle individual differences in EEG signals and improve cross-subject EEG emotion recognition performance\cite{Lan2018DomainAdaptation}\cite{Wu2020TransferLearningReview}. These methods minimize data distribution discrepancies between individuals in the test set (target domain) and those in the training set (source domain). Zheng et al. demonstrated that the transductive parameter transfer (TPT) method could enhance cross-subject emotion recognition accuracy by 19.6\% over the baseline\cite{Zheng2016PersonalizingEEG}. Li et al. proposed a multi-source domain transfer learning method using style transfer mapping to reduce discrepancies between individuals in training and test sets and selecting individuals similar to the test set for model training\cite{Li2019bMultisourceTransferLearning}. Domain adversarial neural networks (DANN)\cite{Ganin2016DomainAdversarialTraining} use a domain discriminator to classify which domain the sample comes from. A gradient reversal layer is inserted between the backbone and the domain discriminator to make the backbone outputs similar across different domains. Combining DANN with fully connected networks, LSTM, or GNN has yielded good performance in cross-subject emotion recognition\cite{Zhong2020EEGEmotionRecognition}\cite{Du2022EfficientLSTM}\cite{LiY2021BiHemisphericDiscrepancy}\cite{Li2019aDomainAdaptation}. 

However, domain adaptation methods still require test-set data to train the domain discriminator, preventing direct generalization of models to test-set individuals. Domain generalization methods\cite{Ma2019AdversarialDomain} have been proposed to address this by using domain adversarial training to learn frequency domain representations shared among training-set individuals, enabling direct application of trained models to test set individuals. Furthermore, researchers have suggested fine-tuning the model with minimal test-set data to enhance performance\cite{Zhao2021PlugAndPlay}. More recently, we propose a contrastive learning method for inter-subject alignment, treating sample pairs from different individuals under similar states as positive samples to extract generalized representations\cite{shen2022contrastive}. Additionally, Wang et al. introduced the denoising mixed mutual reconstruction method for pretraining models\cite{wang2024dmmr}, allowing generalization to new individuals. 

These approaches mitigate individual differences in EEG signals and facilitate cross-subject emotion recognition. However, further exploration into the shared spatiotemporal dynamic patterns of EEG across subjects could potentially enhance performance and deepen our understanding of the mechanisms underlying emotion processing.

\section{Methods}

\begin{figure*}[!t]
\centering
\includegraphics[width=0.98\textwidth]{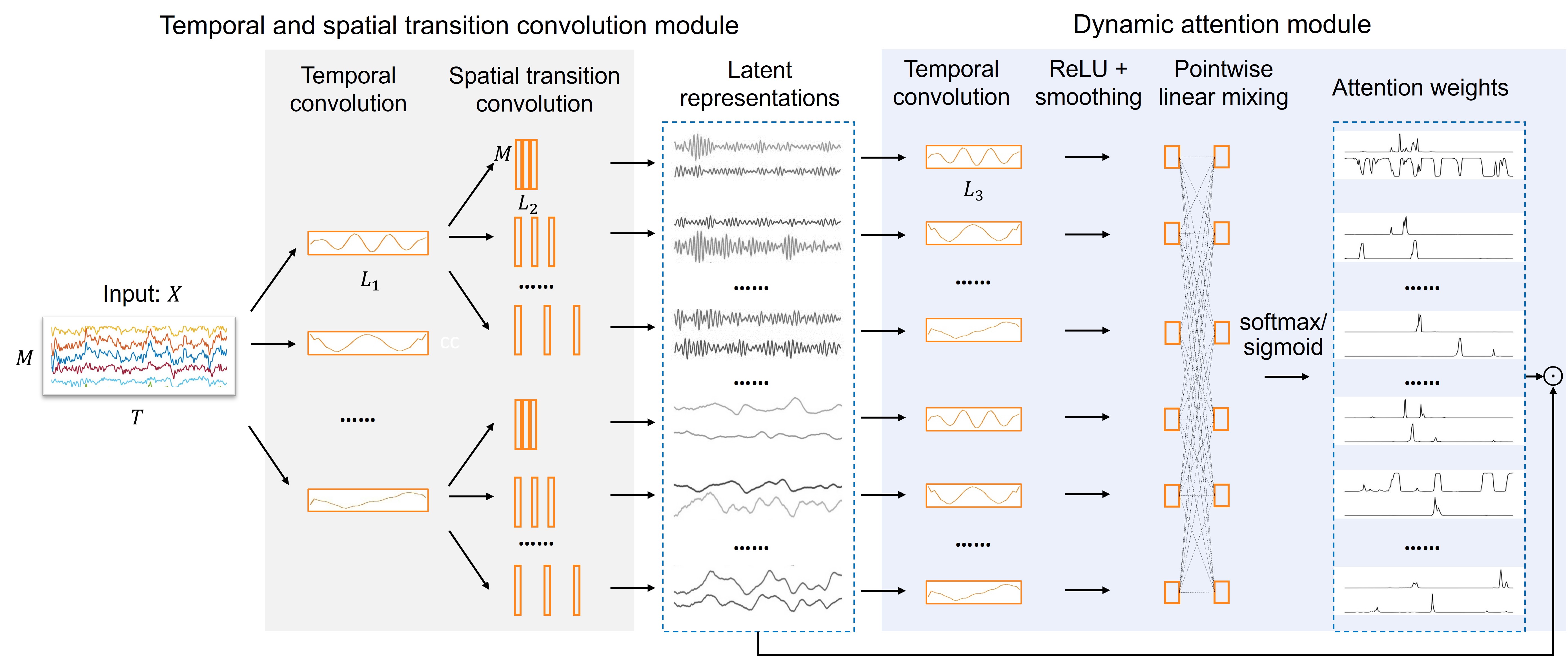}
\caption{The architecture of the dynamic-attention-based EEG state transition model.}
\label{model architecture}
\end{figure*}

The DAEST model is designed to extract the spatiotemporal dynamics of EEG effectively. 
It comprises two main modules: a temporal and spatial transition convolution (TSTC) module and a dynamic attention (DyA) module. To capture temporal-frequency information and rapid spatial variations within EEG, the TSTC module consists of a temporal convolution layer and a spatial transition convolution layer. To select activated dimensions dynamically across different states, the DyA module estimates dynamic attention weights to the spatiotemporal components extracted by the TSTC module. The complete encoder architecture is illustrated in Fig.~\ref{model architecture}. To facilitate cross-subject emotion recognition, a contrastive learning framework is implemented to learn common representations across individuals\cite{shen2022contrastive}. The DAEST model serves as a base encoder in contrastive learning and is trained alongside a subsequent projector. Then the latent representations extracted by the base encoder are used for emotion classification.

\subsection{Temporal and Spatial Transition Convolution (TSTC)}

For EEG inputs $\mathbf{X} \in \mathbb{R}^{M \times  T}$ with M channels and T time points, the encoder first applies one-dimensional temporal convolutional kernels of size $1 \times L_1$ to extract temporal patterns, learning temporal filters to decompose the EEG into different frequency bands. The weights in temporal convolution is denoted as $\mathbf{W}^{temp1} \in \mathbb{R}^{K_1 \times 1 \times L_1}$. The output of the temporal convolution is $\mathbf{H}^{(1)} \in \mathbb{R}^{K_1 \times M \times T}$. $K_1$ is the number of temporal convolutional kernels. The input is padded on the time dimension to make the time points equal to $T$ in the output.

Following this, a spatial transition convolution is employed, utilizing convolutional kernels of size $M \times L_2$ to capture spatial variation patterns across multiple time steps $L_2$. The convolutional kernels use dilations with various temporal intervals, enabling the extraction of EEG variation patterns across different time scales. Unlike the EEGNet \cite{lawhern2018eegnet} and the CLISA \cite{shen2022contrastive} architectures, which use a one-dimensional spatial convolution, the proposed spatial transition convolution can capture rapid changes in EEG over multiple time points. The weights in the spatial transition convolution is denoted as $\mathbf{W}^{spat} \in \mathbb{R}^{K_1K_2 \times M \times L_2}$. The output of the spatial transition convolution is $\mathbf{H}^{(2)} \in \mathbb{R}^{K_1K_2 \times 1 \times T}$. Here, we used group convolution with $K_2$ spatial transition convolutional kernels for each output dimension of the temporal convolution, resulting in $K_1K_2$ dimensions in total. We implemented spatial transition convolution with four different dilations, with $K_1K_2/4$ kernels for each dilation. The hyperparameters are detailed in Section 4.3. Each output dimension of the spatial transition convolution can be regarded as the activation of a specific spatial transition pattern in a specific frequency band. $\mathbf{H}^{(2)}$ can be squeezed into the shape ${K_1K_2 \times T}$ and we denote it as $\mathbf{X}^{latent}$.

\subsection{Dynamic Attention}
The DyA module is applied to the output of the TSTC to estimate time-varying weights for each dimension. Initially, a one-dimensional temporal convolution followed by an average pooling layer along the temporal dimension extracts variation patterns of each dimension over time:
\begin{equation}
\mathbf{A}_{k\cdot} = \mathbf{W}^{\text{temp2}}_{k\cdot} \ast \mathbf{X}^{\text{latent}}_{k\cdot}, k = 1, 2, \ldots, K
\end{equation}
\begin{equation}
\mathbf{\overline{A}}_{k\cdot} = \text{MovingAverage}(\mathbf{A}_{k\cdot})
\end{equation}

\noindent where $K=K_1K_2$ and $\mathbf{W}^{\text{temp2}} \in \mathbb{R}^{K \times L_3}$ are the temporal convolution weights, each row representing a convolution kernel of size $1 \times L_3$ for each dimension of $\mathbf{X}^{latent}$. "$*$" denotes the convolution operation. $\text{MovingAverage}()$ denotes the operation of one-dimensional average pooling.

A pointwise linear mixing layer then performs a linear combination of the dimensions:
\begin{equation}
\mathbf{S}_{i\cdot} = \sum_{k=1}^K \beta_k \mathbf{\overline{A}}_{k\cdot}, \quad i = 1, 2, \ldots, K
\end{equation}

\noindent where $\mathbf{S} \in \mathbb{R}^{K \times T}$. The pointwise linear mixing is also called pointwise convolution, with a convolution kernel of size $1\times1$. These two convolution layers form a depthwise convolution, but with an average pooling layer inserted between them, distinguishing it from a standard depthwise convolution. $\mathbf{S}$ was submitted to an activation function $\phi$ to obtain the attention weights for each dimension at each time point. $\phi$ can be a function with saturation and non-negative characteristics, such as Sigmoid or Softmax:

\begin{equation}
\phi_{sigmoid}(\mathbf{S}_{it}) = \frac{1}{1 + e^{-\mathbf{S}_{it}}}
\end{equation}
\begin{equation}
\phi_{softmax}(\mathbf{S}_{it}) = \frac{e^{\mathbf{S}_{it}}}{\sum_{j=1}^{n} e^{\mathbf{S}_{jt}}}
\end{equation}

\noindent For the Sigmoid function, the attention weights were rescaled to between 0 and 1. For the Softmax function, the attention weights competed against each other. We compared the performance of Sigmoid function and Softmax function in the experiments.

The attention weights are then multiplied with the original output of the TSTC to obtain the weighted activation $\tilde{\mathbf{X}}^{latent} \in \mathbb{R}^{K \times T}$:
\begin{equation}
\tilde{\mathbf{X}}^{latent} = \mathbf{\phi(S)} \odot \mathbf{X}^{latent}
\end{equation}

\noindent $\tilde{\mathbf{X}}^{latent}$ is the output of the base encoder and is submitted to the projector in the contrastive learning procedure. After contrastive learning, $\tilde{\mathbf{X}}^{latent}$ was submitted to the classifier for emotion recognition.

\subsection{The Contrastive Learning Framework}
To address EEG individual differences and improve cross-subject decoding performance, a contrastive learning framework is adopted here\cite{shen2022contrastive}. This framework includes two procedures: a contrastive learning procedure and an emotion classification procedure. The contrastive learning procedure aims to minimize individual differences by maximizing the similarity of EEG representations in similar states across subjects. In the emotion classification procedure, the base encoder trained during the contrastive learning procedure extracts latent representations $\tilde{\mathbf{X}}^{latent}$, and then an emotion classifier is trained for classification.

During the contrastive learning procedure, one EEG sample is selected from each trial of two different individuals per batch, resulting in a batch size of $2N$, where $N$ is the number of trials. Samples corresponding to the same video segment are treated as positive pairs and those from different segments are considered negative pairs. Each sample is sequentially processed by a base encoder and a projector. The projector further extracts temporal variation patterns of EEG using average pooling over the temporal dimension and two one-dimensional temporal convolutional layers. Group convolution with ReLU activations is utilized, with the number of convolution kernels doubled after each layer. The output of the projector is used to compute the normalized temperature-scaled cross-entropy loss, which aims to maximize the cosine similarity of positive pairs relative to negative pairs. 

In the emotion classification procedure, the parameters of the base encoder are fixed, and the latent representations $\tilde{\mathbf{X}}^{latent}$ within each second are averaged over time, resulting in K dimensional features per second. These features are submitted to adaptive feature normalization across time for each individual and smoothed within each trial using a linear dynamical system. The smoothed features are then input into a three-layer Multilayer Perceptron, which outputs the corresponding emotion labels. The readers can refer to our previous study for more details\cite{shen2022contrastive}.

\section{Experiments}
\subsection{Datasets}
We validated our methods on three datasets, FACED, SEED, and SEED-V. SEED is a widely adopted benchmark for emotion recognition. FACED and SEED-V are more recently published datasets containing more emotion categories, with FACED featuring a much larger number of subjects than other publicly available datasets.

FACED is an affective computing EEG dataset with the largest number of subjects (N = 123), which provides a good benchmark for cross-subject emotion recognition methods. The dataset includes nine emotion categories elicited through film clips: anger, fear, disgust, sadness, amusement, inspiration, joy, tenderness, and neutral. Each participant underwent 28 trials—four for the neutral emotion and three for each of the other emotions. Each trial consisted of four stages: fixation, film clip presentation, self-report ratings, and rest. The film clips have a duration of 34 to 129 seconds. During emotion elicitation, 32-channel EEG signals were recorded using the NeuSen.W32 system (Neuracle, China). We conducted two types of classifications on the FACED dataset: a nine-class classification of fine-grained discrete emotions and a binary classification distinguishing positive emotions (including amusement, inspiration, joy, and tenderness) from negative emotions (including anger, fear, disgust, and sadness). 

SEED provides 62-channel EEG recordings (ESI NeuroScan System) from 15 subjects during emotion elicitation. Positive, neutral, and negative emotions were induced by film clips. The experiment included 15 trials per session, with three sessions conducted in total, spaced one-week or longer apart. Each trial consisted of a start cue, movie clip presentation, self-assessment, and rest. There were five trials for each emotion. The movie clips have a duration ranging from 185 to 265 seconds. We performed a three-class classification on the SEED dataset. 

SEED-V is an extension of the SEED dataset, with five emotion categories elicited: happiness, sadness, disgust, fear, and neutral. It includes 62-channel EEG recordings (ESI NeuroScan System) from 20 subjects during emotion elicitation. In each session, 15 film clips were used, with three clips for each emotion. Each participant completed three sessions. The experimental procedure closely mirror that of the SEED dataset, consisting of a start cue, movie clip presentation, self-assessment, and rest for each trial. We performed a five-class classification on the SEED-V dataset.

\subsection{Data Preprocessing}

To ensure consistency and comparability across the three datasets, we implemented a standardized automatic preprocessing pipeline using Matlab. First, EEG signals were downsampled to 125 Hz and filtered with a 0.5-47 Hz bandpass filter. The signals were then segmented into trials based on the onset and offset of emotional video stimuli. For each channel, if the proportion of data exceeding a specified multiple (m) of the median value was greater than a certain percentage (n) of the trial duration, the channel was classified noisy. We used two sets of thresholds: m=3, n=0.4 to identify long-lasting artifacts, and m=30, n=0.01 to detect short-term large artifacts. Noisy channels identified in this manner were interpolated using their three nearest neighboring channels. Next, independent Component Analysis (ICA) was applied to remove artifacts caused by eye movements or muscle activity. We utilized automatic component labeling tools in EEGLab. Components labeled as eye- or muscle-related with a confidence level exceeding 0.8, the default thresholds in EEGLab, were removed. On average, 5.4 out of 30 components were labeled as noise for the FACED dataset, 8.9 out of 60 for SEED, and 12.4 out of 60 for SEED-V.

During the initial noisy channel detection, we observed that frontal channels (such as Fp1 and Fp2) were frequently identified as noisy, which interfered with the subsequent ICA-based detection of eye movement components. To address this issue, we excluded channels Fp1, Fp2, F7, and F8 from the initial noisy channel interpolation to preserve eye movement features. After applying ICA, the noisy channel interpolation again, this time including all channels. Finally, the data were re-referenced to the common average.

\subsection{Implementation Details}
This section provides details on model hyperparameter settings and training procedures. Unless otherwise stated, all settings refer to those used on the FACED dataset. Any different settings for SEED and SEED-V will be explicitly noted.

In the TSTC module, the temporal convolution utilizes $K_1=16$ one-dimensional filters, each with a length of $L_1=30$, translating to an actual time duration of 240 ms ($1000\cdot L_1/f_s$, where $f_s=125$). There are $K_2=16$ spatial transition convolution kernels for each temporal convolution kernel, resulting in $K_1*K_2=256$ spatial transition convolution kernels in total. The spatial transition convolution covers $L_2=3$ steps in the temporal dimension. There are four sets of two-dimensional dilated convolution kernels, with dilations of 1, 3, 6, and 12 on the temporal dimension, respectively. Each of the four sets contains $K_1K_2/4=16*16/4=64$ convolution kernels. In the DyA module, the temporal convolution also employs group convolution with a kernel length of $L_3=7$. The subsequent average pooling has a window length of $L_3=7$ and a stride of 1. In The projector, the average pooling window length and stride are all set to 15 and the kernel sizes of the two convolutional layers are both set to 3. The emotion classifier is a multilayer perceptron with two hidden layers of 128 neurons and 64 neurons, respectively, with ReLU activation between layers. 


For model training in the contrastive learning procedure, we set the learning rate as $0.0007$ and the weight decay as $0.00015$. The model was trained for 30 epochs with an early stopping patience of 10 epochs. The time length of the samples in contrastive learning was determined by a tradeoff between the training samples’ number and adequate sample length. We set the average sample number for each video to 13 in the three datasets. The sample number is formulated as $N_{sample} = \langle \lfloor(timeLen_{v_i}-timeLen_{sample})/stepLen\rfloor \rangle_i +1$, where $timeLen_{v_i}$ is time length of video $i$, $timeLen_{sample}$ and $stepLen$ are individual sample length  and sampling interval, where $stepLen = timeLen_{sample} // 2$. The sample lengths for FACED, SEED, and SEED-V datasets are 5, 30, and 22 seconds, respectively. In the emotion classification procedure, we set the learning rate as $0.0005$. The weight decay was selected from 0.001, 0.0022, 0.005, 0.011, and 0.025 by cross-validation. The classifier was trained for 100 epochs with an early stopping patience of 30 epochs.

\subsection{Performance Comparison}

We compared the performance of the proposed DAEST model with state-of-the-art domain generalization models on the three datasets, including CLISA\cite{shen2022contrastive}, GCPL\cite{li2024generalized}, DResNet\cite{ma2019reducing}, PPDA\cite{zhao2021plug}, LDG+Resnet101\cite{tao2023local}, PCDG\cite{cai2023two}, and DMMR\cite{wang2024dmmr}. Additionally, we evaluated the proposed model against variations in network architecture to assess the effectiveness of dynamic attention mechanisms. These variations included a model without attention, one with global channel attention, and another using Transformer layers. In the model without attention, the DyA module was removed, while all other settings remained identical to the complete model. For the global channel attention model, the average pooling size at the DyA module's output was equal to the number of time steps in the sample, thus the attention weights were constant for each channel within an EEG sample. Other settings were the same as the proposed model. For the Transformer-based models, we replaced the DyA module with two types of Transformer layers, one applying self-attention across the feature dimension to learn interactions among different latent dimensions, and another applying self-attention across the temporal dimension to capture potential variation patterns in time sequences. Besides, a baseline model (DE+MLP) without contrastive learning was implemented, in which the DE features were directly extracted from the EEG signals and submitted to adaptive normalization, smoothing, and emotion classification. We also compared the dynamic attention module's performance using different activation functions, including Sigmoid, Softmax, and ReLU. To validate the TSTC module's design, we evaluated lesioned models with no temporal convolution, no spatial transition convolution, and no dilations in spatial transition convolution. 

\section{Results}

\subsection{Comparison with State-of-the-Art Methods on Three Datasets}

The comparison of cross-subject emotion recognition performance between the DAEST model and other state-of-the-art methods on the FACED, SEED, and SEED-V datasets are presented in Tables~\ref{table1} and~\ref{table2}.

\begin{table}[!t]
\renewcommand{\arraystretch}{1.3}
\caption{Cross-subject emotion recognition performance on the FACED dataset}
\label{table1}
\centering
\begin{tabular}{@{}lccc@{}}
\hline
Methods    & FACED-2 (Acc\%) & FACED-9 (Acc\%) &  \\ \hline
DE+MLP     & 60.9$\pm$3.2    &   35.0$\pm$4.6  \\ 
CLISA\cite{shen2022contrastive}     &  67.8$\pm$4.1      &   43.2$\pm$5.9  \\ 
GCPL\cite{li2024generalized}    &    /   &   36.9$\pm$3.3  \\ 
\textbf{DAEST (ours)} &  \textbf{75.4$\pm$5.5}   &  \textbf{59.3$\pm$7.7}   \\ \hline
\end{tabular}
\end{table}


\begin{table}[!t]
\renewcommand{\arraystretch}{1.3}
\caption{Cross-subject emotion recognition performance on the SEED and SEED-V dataset}
\label{table2}
\centering
\begin{tabular}{@{}lcc@{}}
\hline
Methods    & SEED (Acc\%) & SEED-V (Acc\%) \\ \hline
DE+MLP     & 79.9$\pm$8.7    &   59.3$\pm$17.2   \\ 
DResNet\cite{ma2019reducing}     & 85.3$\pm$8.0    &   /   \\
PPDA \cite{zhao2021plug}       & 86.7$\pm$7.1    &   /   \\
CLISA\cite{shen2022contrastive}       & 86.4$\pm$6.4     &  67.3$\pm$13.0    \\
LDG+Resnet101\cite{tao2023local}   &  82.0$\pm$5.9 &  /  \\
PCDG\cite{cai2023two}  & 87.3$\pm$2.1     &  /  \\
GCPL\cite{li2024generalized}   & 80.7$\pm$6.0    &   /  \\ 
DMMR\cite{wang2024dmmr}  & \textbf{88.3$\pm$5.6}   &   /  \\
\textbf{DAEST (ours)} &   \textbf{88.1$\pm$3.6} &  \textbf{73.6$\pm$12.7}    \\ \hline
\end{tabular}
\end{table}

On the FACED dataset, the DAEST model achieved an accuracy of 75.4$\pm$5.5\% for binary classification (FACED-2) and 59.3$\pm$7.7\% for nine-class classification (FACED-9)(Table~\ref{table1}). In comparison, existing domain generalization methods yielded much lower accuracies: the DE+MLP model achieved 60.9$\pm$3.2\% and 35.0$\pm$4.6\% for the binary and nine-class tasks, respectively; the CLISA model reached 67.8$\pm$4.1\% and 43.2$\pm$5.9\%; and the GCPL model achieved an accuracy of 36.9$\pm$3.3\% for the nine-class task. These results indicate that the DAEST model dramatically outperforms other methods on the FACED dataset, particularly in the nine-class task, where the accuracy increases by 22.4\% over GCPL and 16.1\% over CLISA, underscoring its superiority in handling more fine-grained emotion classification tasks.

On the SEED and SEED-V datasets, the DAEST model also demonstrated exceptional cross-subject emotion recognition capability (Table~\ref{table2}). It achieved an accuracy of 88.1$\pm$3.6\% on the SEED dataset, comparable to the state-of-the-art methods. On the SEED-V dataset, the DAEST model achieved an accuracy of 73.6$\pm$12.7\%, outperforming other domain generalization methods. In comparison, the DE+MLP model achieved accuracies of 79.9$\pm$8.7\% and 59.3$\pm$17.2\% on the SEED and SEED-V datasets, respectively, while the CLISA model achieved accuracies of 86.4$\pm$6.4\% and 67.3$\pm$13.0\%. Consistent with the results on the FACED dataset, the DAEST model exhibited a larger performance improvement on the more fine-grained emotion classification task on SEED-V, with a 6.3\% increase in accuracy.

\subsection{Results of Model Variants and Ablation Studies}

To evaluate the effectiveness of each module in the DAEST model, we tested various model variants and performed ablation studies. The results, summarized in Table~\ref{table3}, show the impact of each module on emotion recognition performance.

\begin{table}[!t]
\renewcommand{\arraystretch}{1.3}
\caption{Performance comparison of different model architectures}
\label{table3}
\centering
\resizebox{\columnwidth}{!}{ 
\begin{tabular}{lcc}
\hline
Methods & FACED-2 (Acc\%) & FACED-9 (Acc\%)\\ 
\hline
\multicolumn{3}{l}{DyA module} \\ 
\hline
Without attention & 72.9$\pm$4.6 & 47.6$\pm$7.4 \\ 
Global channel attention & 72.0$\pm$4.4 & 52.6$\pm$8.9 \\ 
Transformer layers (across features) & 64.5$\pm$5.1 & 33.7$\pm$5.8 \\ 
Transformer layers(across time) & 74.3$\pm$4.4 & 55.1$\pm$7.2 \\ 
Dynamic attention (ReLU) & 72.7$\pm$4.0 & 50.9$\pm$8.5 \\ 
Dynamic attention (Softmax) & 74.6$\pm$4.3 & 56.2$\pm$7.2 \\ 
\textbf{Dynamic attention (Sigmoid)} & \textbf{75.4$\pm$5.5} & \textbf{59.3$\pm$7.7} \\ 
\hline
\multicolumn{3}{l}{TSTC module} \\ 
\hline
No temporal convolution & 70.3$\pm$4.4 & 48.3$\pm$9.3 \\ 
No spatial transition convolution & 70.0$\pm$4.0 & 49.3$\pm$6.2 \\ 
No dilation & 71.9$\pm$5.3 & 53.6$\pm$9.0 \\ 
\hline
\end{tabular}
}
\end{table}

In examining the impact of the DyA module on model performance, we found that the dynamic attention module outperformed models without attention, as well as those utilizing global channel attention and various types of Transformer-based (self-attention) models. Compared to the complete model, models without attention showed an accuracy decrease of 2.5\% and 11.7\% on the FACED-2 and FACED-9 datasets, respectively. Global channel attention achieved accuracies of 72.0$\pm$4.4\% and 52.6$\pm$8.9\% on FACED-2 and FACED-9, respectively, which were 3.4\% and 6.7\% lower than the dynamic attention, indicating the importance of dynamically estimating attention weights in emotion recognition. The Transformer-based model across feature dimensions performed the worst, with accuracies of 64.5$\pm$5.1\% and 33.7$\pm$5.8\% on FACED-2 and FACED-9, respectively. The Transformer-based model across time dimensions performed relatively better with 74.3$\pm$4.4\% and 55.1$\pm$7.2\% but still fell short by 1.1\% and 4.2\% compared to the dynamic attention model. This suggests that the design of the dynamic attention module is superior to the state-of-the-art self-attention architectures with fewer parameters (The size of our model is 0.36M and that of the Transformer-based model across time is 2.97M). Among the different activation functions used in the dynamic attention module, the Sigmoid function produced the best results, followed by Softmax, which achieved 74.6$\pm$4.3\% and 56.2$\pm$7.2\% on FACED-2 and FACED-9, respectively. ReLU performed the worst. These findings highlight the importance of scaling the dynamic attention weights within the 0-1 range for optimal model performance.

In evaluating the TSTC module's design on the model's performance, we found that removing the temporal convolution led to a decrease in accuracy by 5.1\% and 11.0\% on the FACED-2 and FACED-9 datasets, respectively. The model without spatial transition convolution exhibited a decrease in accuracy by 5.4\% and 10.0\%, indicating that both temporal and spatial transition convolutions have a significant impact on emotion recognition performance. The removal of dilation in spatial transition convolution resulted in a 3.5\% and 5.7\% decrease in accuracy on the FACED-2 and FACED-9 datasets, respectively, demonstrating the importance of covering multiple time scales with dilations for spatial transition convolution.

\begin{figure*}[!t]
\centering

\subfloat[FACED-9, w/o DyA module]{\includegraphics[width=0.48\textwidth]{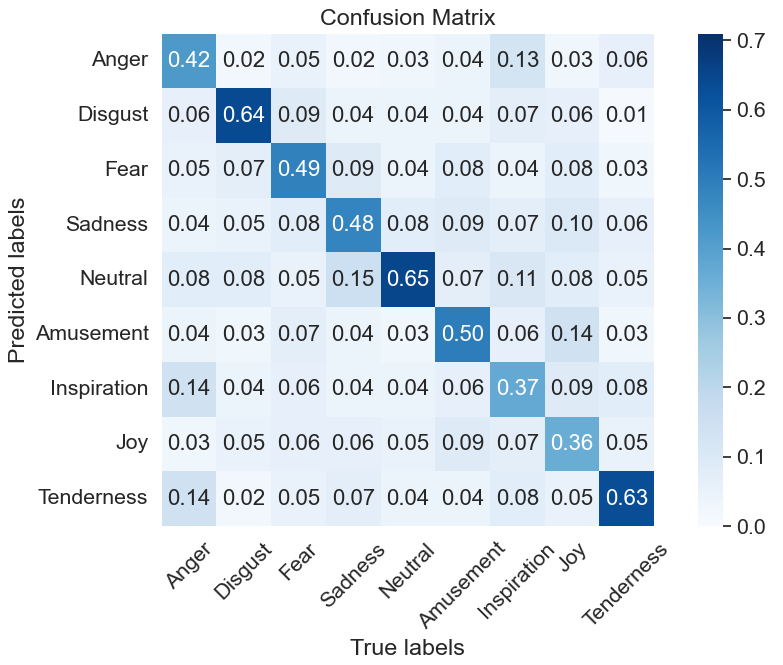}%
\label{fig_faced9_case1}}
\hfil
\subfloat[FACED-9, with DyA module]{\includegraphics[width=0.48\textwidth]{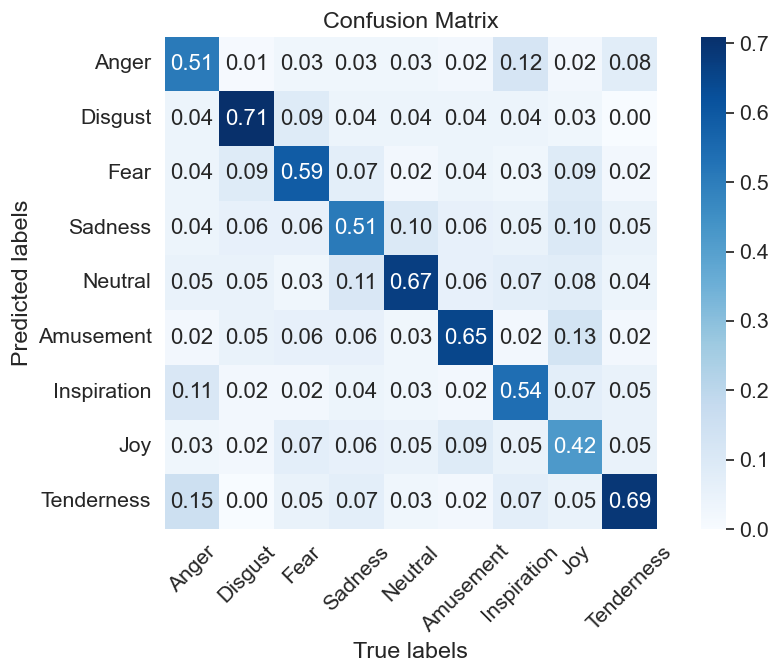}%
\label{fig_faced9_case2}}
\\
\subfloat[SEED, w/o DyA module]{\includegraphics[width=0.22\textwidth]{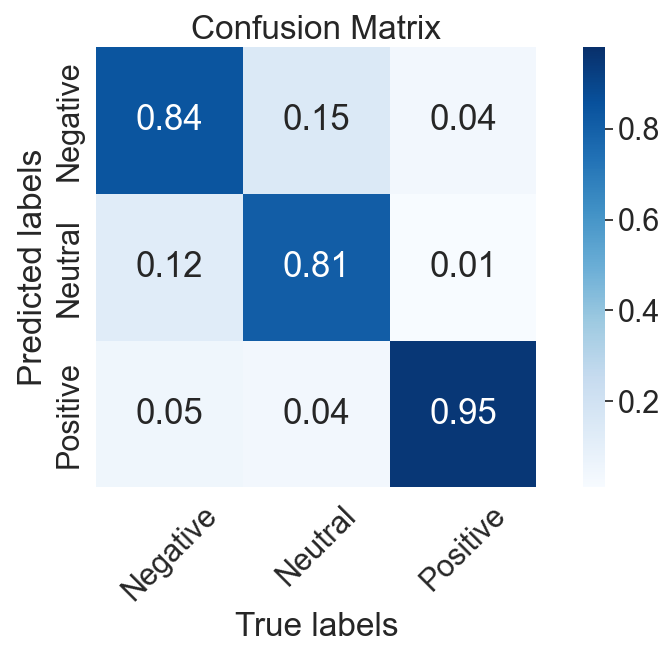}%
\label{fig_seed_case1}}
\hfil
\subfloat[SEED, with DyA module]{\includegraphics[width=0.22\textwidth]{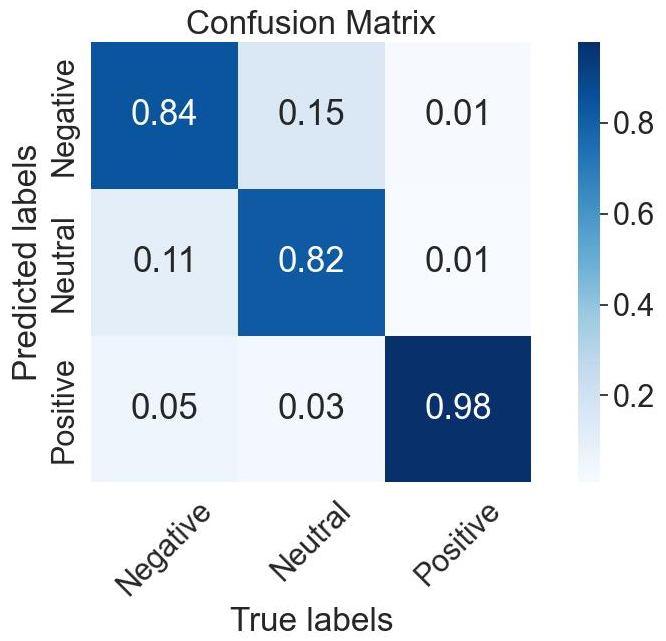}%
\label{fig_seed_case2}}
\hfil
\subfloat[SEED-V, w/o DyA module]{\includegraphics[width=0.25\textwidth]{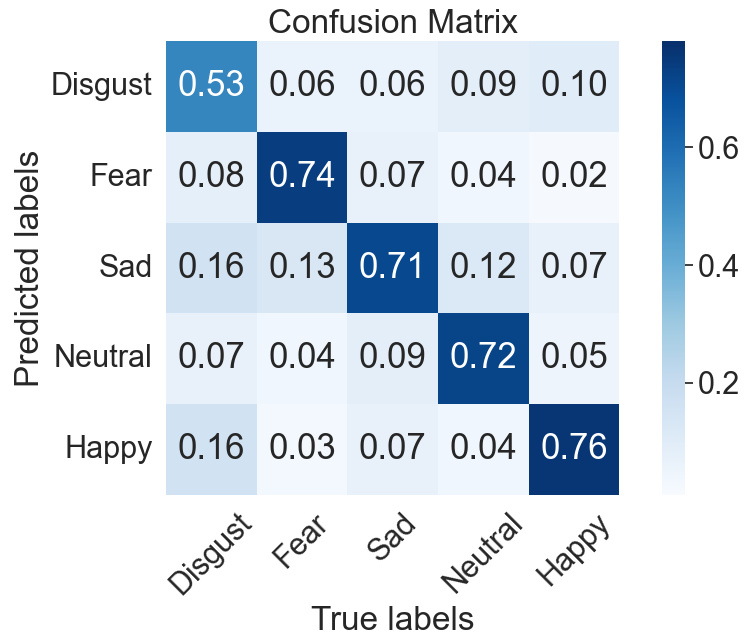}%
\label{fig_seedv_case1}}
\hfil
\subfloat[SEED-V, with DyA module]{\includegraphics[width=0.25\textwidth]{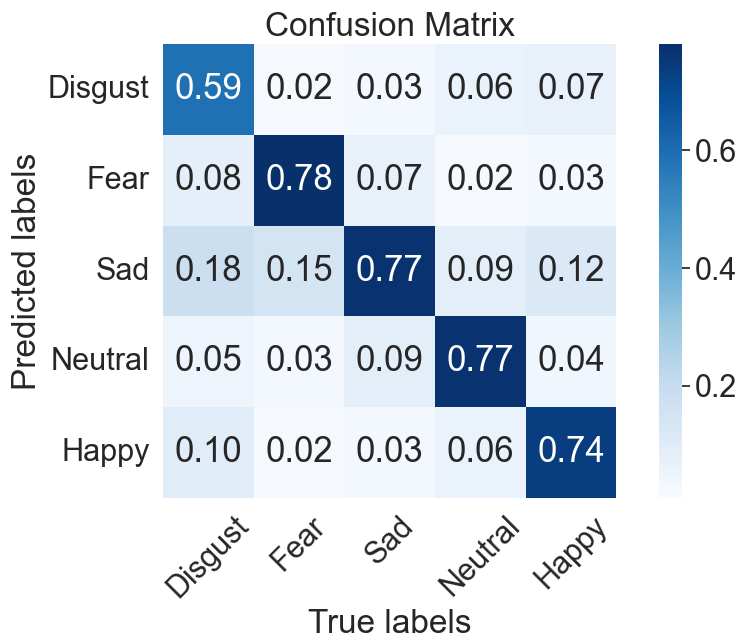}%
\label{fig_seedv_case2}}
\caption{Confusion matrices on the FACED-9, SEED, and SEED-V classification tasks (without DyA module and with DyA module).}
\label{fig2}
\end{figure*}

To further investigate the effects of the dynamic attention mechanism on recognition of each emotion category, we generated confusion matrices for the FACED-9, SEED, and SEED-V datasets (Fig.~\ref{fig2}), comparing models with and without the DyA module. For the FACED-9 dataset, the DyA module led to the greatest improvements in recognizing the positive emotion categories of amusement, inspiration, and joy, with accuracy increases of 15\%, 17\%, and 6\%, respectively. The classification accuracies of negative emotions, such as anger, disgust, and fear were also enhanced by 9\%, 7\%, and 10\%, respectively. This indicates that the dynamic attention mechanism enhances the model's ability to distinguish between nuanced emotions. For the SEED dataset, the DyA module led to improvements in recognizing positive and neutral emotions, with accuracy increases of 3\% and 1\%, respectively. In the SEED-V dataset, the DyA module enhanced the model's ability to recognize nuanced negative emotions, with notable accuracy gains of 6\% for both disgust and sad and 5\% for neutral. Additionally, there was a reduction in the misclassification rates of these negative emotions as positive (i.e., happy), contributing to a more refined emotional recognition overall.

\subsection{Visualization of the Learned EEG Spatiotemporal Dynamics}

\begin{figure*}[!t]
\centering
\includegraphics[width=0.95\textwidth]{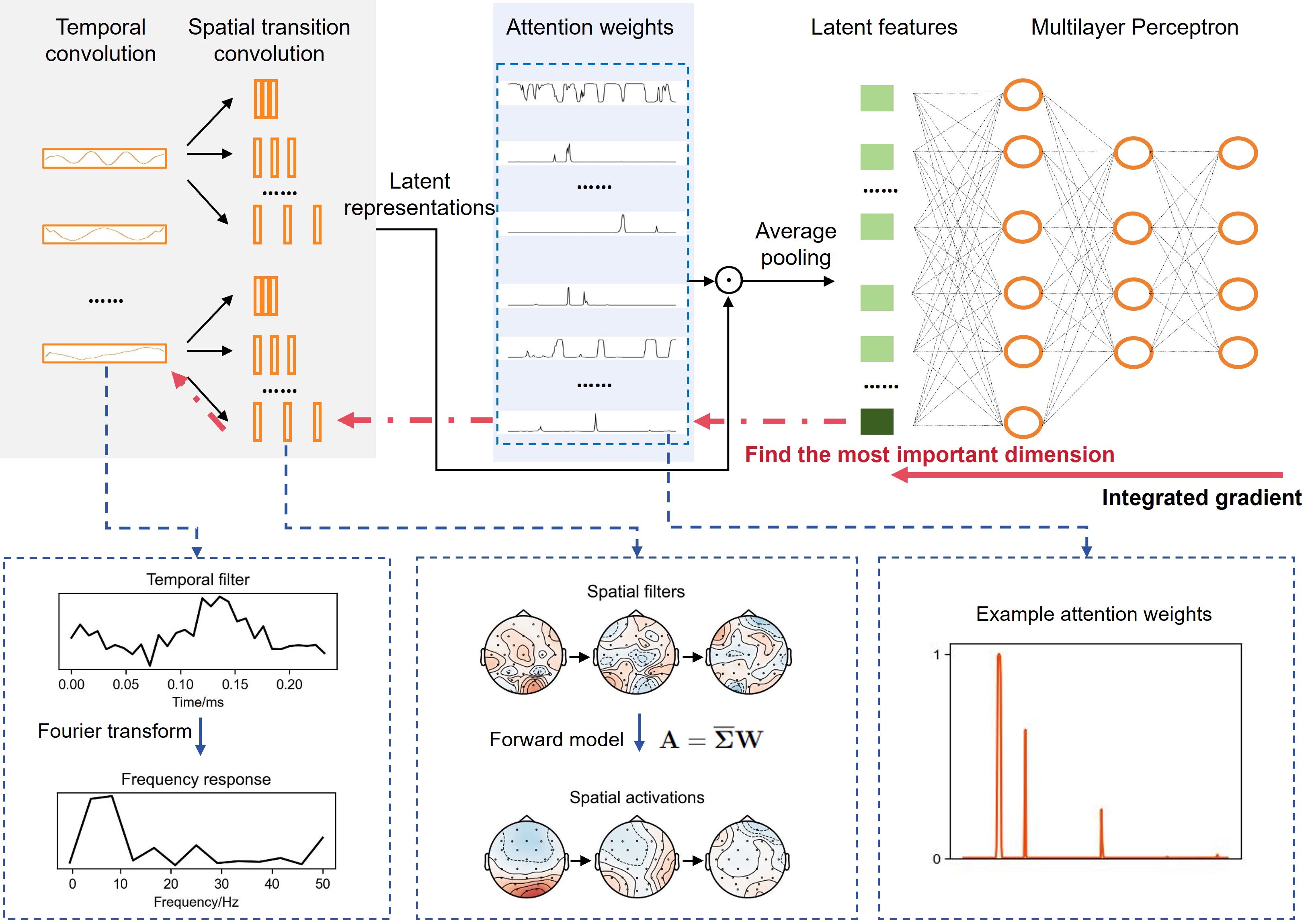}
\caption{The pipeline of interpretability analysis.
The integrated gradient method is used to identify the most important dimension or feature for each emotion in the Multilayer Perceptron. The temporal convolution kernel (or temporal filter), the spatial transition convolution kernel (or spatial filters), and the attention weights that produce this feature are visualized. Fourier transform of the temporal filter is identified as its frequency responses. Spatial activations are defined as multiplications of spatial filters with their corresponding inputs' covariance.}
\label{fig3-0}
\end{figure*}

\begin{figure*}[!t]
\centering
\includegraphics[width=0.95\textwidth]{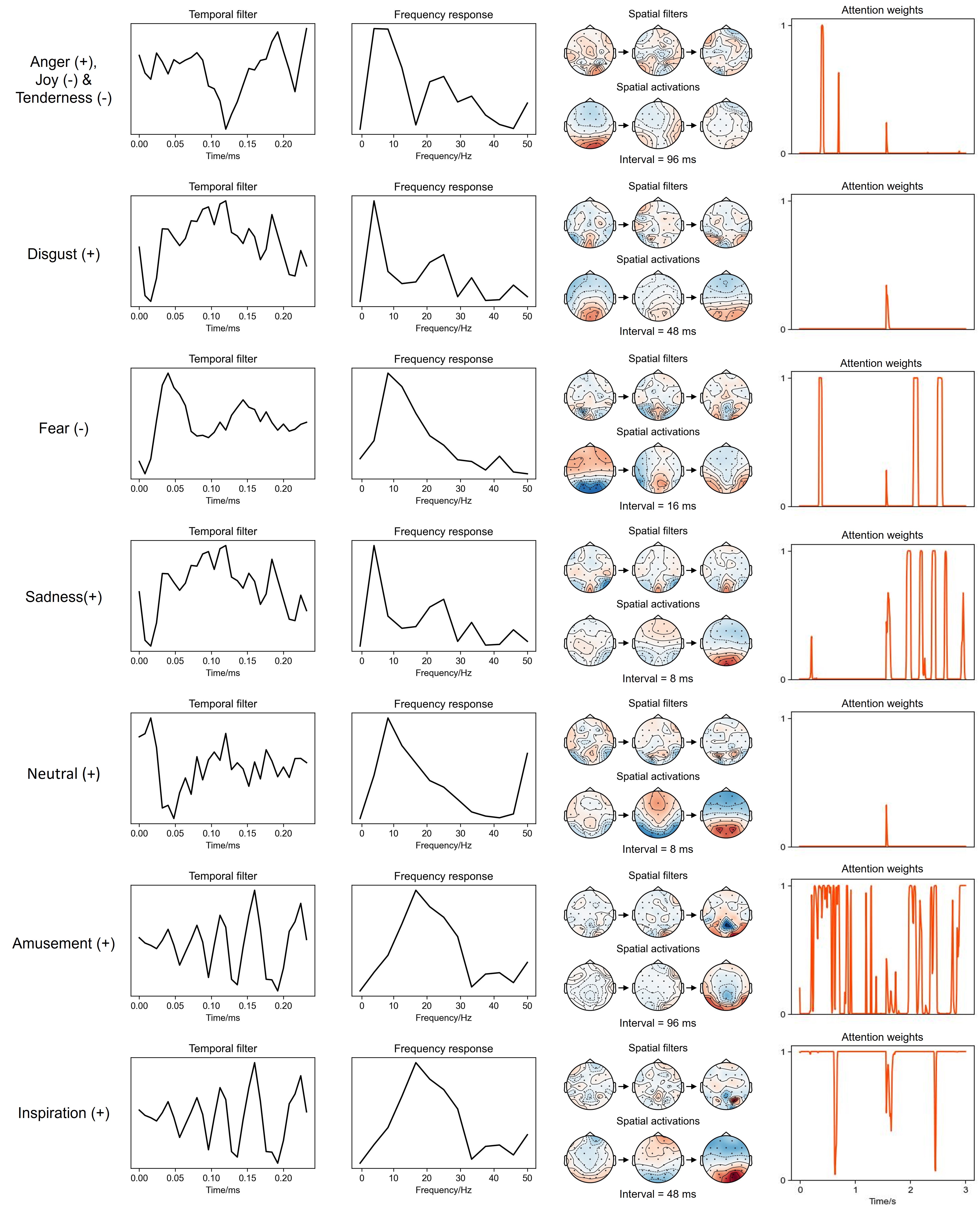}
\caption{Visualization of the spatiotemporal dynamics for the most important dimension of each emotion on the FACED-9 task.
The temporal filters (the first column) and their frequency response (the second column), spatial filters and spatial activations (the third column), and example segments of attention weights (the fourth column) for the most important dimensions are shown. The “+”/”-” symbols in the parentheses following the emotion category name indicate that an increase in that feature corresponds to a high/lower probability of the corresponding emotion. The "interval" noted below spatial activations refers to the time interval of dilations for the corresponding spatial transition convolution kernel.}
\label{fig3}
\end{figure*}

To investigate the important EEG spatiotemporal components and their dynamics underlying emotion processing, we used the integrated gradient method to identify feature contributions to each emotion in the classifier of the FACED-9 task. A feature input to the classifier was extracted from a latent dimension in the encoder. A latent dimension was produced by a temporal convolution, a spatial transition convolution, and multiplied by attention weights in the encoder (Fig.~\ref{fig3-0}). Thus, the temporal-frequency characteristics, spatial transition patterns, and the dynamic activations of a dimension can be directly reflected in the temporal convolution kernels (or temporal filters), spatial transition convolution kernels (or spatial filters), and the dynamic attention weights, respectively. We visualized the temporal filters and their frequency response, the spatial filters and spatial activations, and an example segment of attention weights for the most important dimension of each emotion in (Fig.~\ref{fig3}). The spatial activation is calculated by $\mathbf{A}^{spat}_{k\cdot t}=\overline{\mathbf{\Sigma}}_k\mathbf{W}^{spat}_{k\cdot t}$\cite{haufe2014interpretation}. It transforms the parameters $\mathbf{W}^{spat}$ of a backward projection model to those ($\mathbf{A}^{spat}$) of a forward model. $\overline{\mathbf{\Sigma}}$ is the average covariance matrix across the input dimensions of the convolution kernel $\mathbf{W}^{spat}_{k\cdot\cdot}$.


Three emotions - anger, joy, and tenderness - share the same most important feature, with an increase of this feature leading to a higher probability of anger and a decrease corresponding to a higher probability of joy and tenderness. This feature has the highest frequency response in 4-10 Hz and a spatial activation transitioned from the occipital region to bilateral parietal regions and bilateral frontal regions. This dimension was sparsely activated as shown by the attention weights (the first row in Fig.~\ref{fig3}). Disgust and sadness shared the same frequency response with a peak of 4 Hz, but differed in their spatial activation transitions and attention weight patterns, with the important dimension for sadness activated more frequently. Fear had a peak frequency response at 8-12 Hz, with a spatial activation transitioned from anterior-posterior activation to middle occipital activation and then to bilateral occipital activation. The other two positive emotions, amusement and inspiration had a higher frequency response of 12 to 30 Hz, overlapping with the beta band. These two dimensions have much higher attention weights than other dimensions.  

\begin{figure}[!t]
\centering
\includegraphics[width=0.45\textwidth]{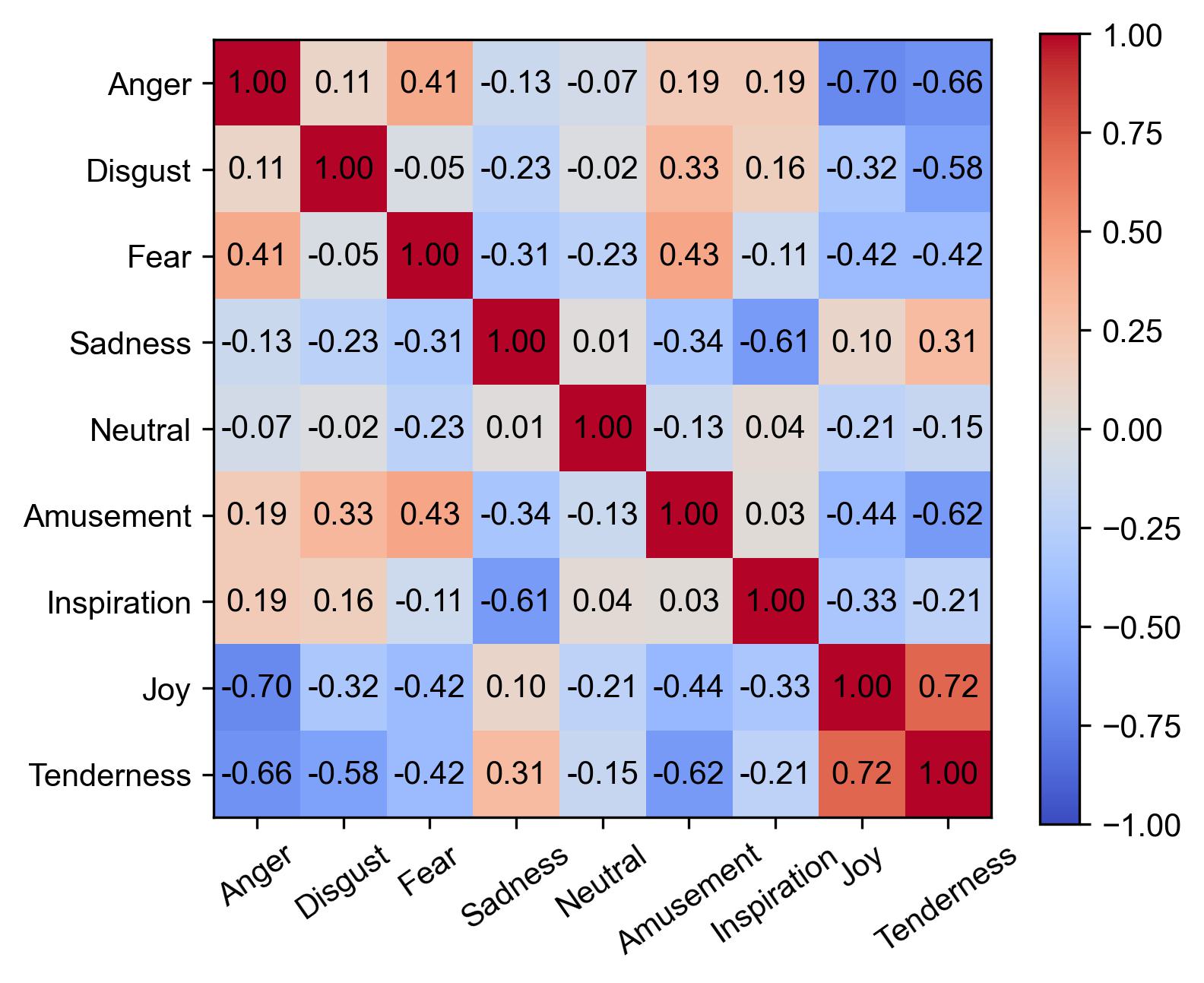}
\caption{Correlation of feature contributions to each emotion on the FACED-9 task.}
\label{fig4}
\end{figure}

To better understand the relationship of feature contributions to different emotions, we calculated the correlation of feature contributions between every two emotions (Fig.~\ref{fig4}). A higher correlation coefficient between two emotions means more overlapping features contribute to both these emotions. Negative correlations indicate features tend to contribute reversely to the two emotions. Joy and tenderness have the most similar feature contributions, with a correlation of 0.724. Among negative emotions, anger and fear have the highest correlation of 0.410. Interestingly, notable correlations were observed between positive emotions and negative emotions. For example, the correlations between amusement and fear, amusement and disgust, and tenderness and sadness all exceeded 0.3. The strongest negative correlations were observed between joy and anger (-0.704), tenderness and anger (-0.660), tenderness and amusement (-0.624), and inspiration and sadness (-0.611). These findings suggest that the underlying neural activities associated with different emotions share common features to some extent and that the grouping of emotions may not be strictly based on negative and positive valence.

\subsection{The Effects of Hyperparameters and Data Noise}

We investigated the effects of two important hyperparameters on model performance, namely the time window of dynamic attention and the time steps of spatial transition convolution, on the FACED dataset. 

\begin{figure*}[!t]
\centering
\subfloat{\includegraphics[width=\textwidth]{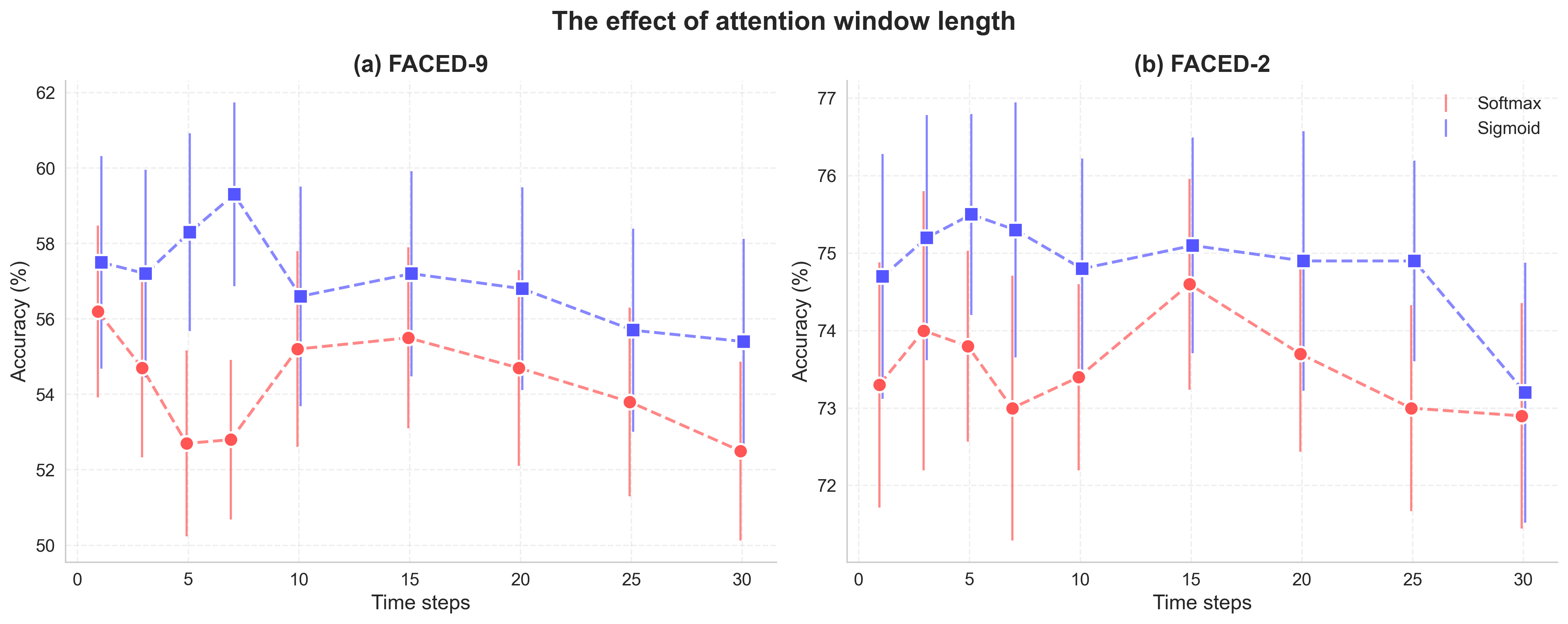}%
\label{fig_first_case}}
\hfil
\subfloat{\includegraphics[width=\textwidth]{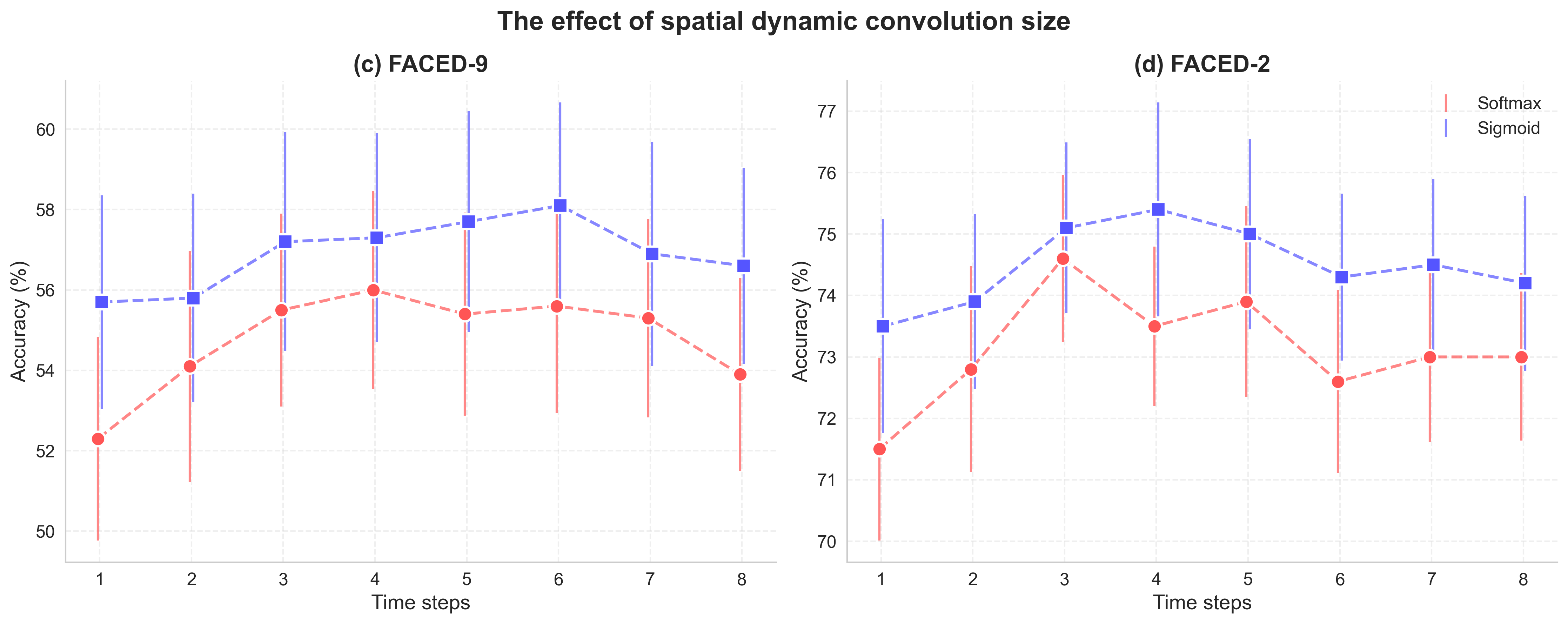}%
\label{fig_second_case}}
\caption{The performance with varied hyperparameters (on the FACED task).}
\label{fig_sim}
\end{figure*}

We evaluated the effect of attention window length (\(L_3\)) on model performance by varying its values from 1 to 30 while keeping other parameters constant. The results showed that performance generally peaked within the ranges of 1-10 and 10-20 (Fig.~\ref{fig_sim}a,b). For the FACED-9 task, the sigmoid model achieved optimal performance at \(L_3=7\) with an accuracy of $59.3\pm7.7$, while the softmax model peaked at \(L_3=1\) with an accuracy of $56.2\pm 7.2$. In the FACED-2 task, the sigmoid model performed best at \(L_3=5\) with an accuracy of $75.4\pm5.5$, and the softmax model peaked at \(L_3=15\) with an accuracy of $74.6\pm 4.3$. Considering the performance of both FACED-2 and FACED-9 tasks, \(L_3=15\) is identified as a generally effective parameter setting.

We also examined how varying the time steps of the spatial transition convolution (\(L_2\)) from 1 to 8 affected model performance while keeping other parameters constant (Fig.~\ref{fig_sim}c,d). Increasing \(L_2\) to 3 resulted in a dramatic improvement in performance. The performance continued to improve until \(L_2\) reached 6, after which no further gains were observed. Thus, \(L_2 = 3\) is identified as the optimal setting, balancing model efficiency and performance. The peak performance varied slightly with different activation functions and classification tasks, with sigmoid generally outperforming softmax.

We also experimented to see the effect of data noise on model performance. The complete preprocessing used the default threshold in EEGLab to remove eye and muscle components in ICA. Here, we test the effect of less noise removal on the emotion recognition performance. Specifically, we only remove the eye-blink components in ICA. Surprisingly, the accuracy increased dramatically by 4.8\% for the binary classification of the FACED dataset and by 8.0\% for the nine-class classification of the FACED dataset. The results of less noise removal are comparable to those obtained using the officially preprocessed data, where primarily eye-blink components were removed. The results suggest that activities typically considered as EEG noise might contain useful information for emotion recognition. Since this study focuses on the emotion recognition capability of brain components and aims to provide insights into the brain mechanisms underlying emotion processing, we used data from the complete preprocessing pipeline throughout our analysis. Future studies could explore the effects of isolating eye-related and muscle-related noise on emotion recognition.

\begin{table}[!t]
\renewcommand{\arraystretch}{1.3}
\caption{The effect of denoising on the performance}
\label{table_example}
\centering
\resizebox{\columnwidth}{!}{ 
\begin{tabular}{@{}lccc@{}}
\hline
Methods    & FACED-2 (Acc\%) & FACED-9 (Acc\%) &  \\ \hline
Complete preprocessing     &  75.4$\pm$5.5 & 59.3$\pm$7.7 \\  
Only removing eye-blink component in ICA    & 80.2$\pm$4.9    &   67.3$\pm$8.8  \\ 
Official preprocessing   & 81.7$\pm$4.3    &   67.9$\pm$7.3  \\ 
\hline
\end{tabular}
} 
\end{table}

\section{Discussion}
This study proposes a biologically interpretable dynamic attention mechanism to model the state transitions in EEG activities. The dynamic attention mechanism has been shown to significantly enhance the performance of emotion recognition across three datasets, underscoring the critical role of neural state transitions underlying emotion processing. In addition, we extend the spatial convolution to cover more time steps with varying dilation sizes, which effectively captures spatial transition patterns at multiple time scales. These architectural designs enable the network to comprehensively learn the spatiotemporal dynamics within EEG signals. 

In model comparisons, the performance enhancement of the DAEST model in more fine-grained discrete emotion classification is notable, with an accuracy increase of 16.1\% on the FACED-9 task and an accuracy increase of 6.3\% on the SEED-V dataset in comparison to the state-of-the-art method. The higher performance of DAEST over CLISA (contrastive learning with only temporal and spatial convolutions in the encoder) indicates the effectiveness of spatial transition convolution and dynamic attention designs for discrete emotion classification. From the confusion matrices, we can see that the more challenging discrimination of distinct emotions with the same valence is enhanced by dynamic attention mechanisms, especially amusement, inspiration, and joy. The dynamic attention mechanisms also significantly outperformed global channel attention, which estimates a constant attention weight for each dimension. Similar designs have been adopted in previous studies\cite{Tao2020ChannelWiseAttention}. This indicates the importance of considering dynamic weights on the latent dimensions, which model the change of neural states. In summary, DAEST is a lightweight interpretable model for EEG-based emotion recognition with an extraordinary ability in fine-grained discrete emotion recognition.


The spatiotemporal dynamic patterns learned by the model shed light on neural mechanisms of discrete emotion processing. Anger, joy, and, tenderness share the most important feature and lie on two poles of this attribute, i.e., the increase of this feature is associated with more anger, and the decrease of this feature is associated with more joy or tenderness. Other discrete emotions all have distinct important spatiotemporal dynamic patterns. The positive emotions, amusement, and inspiration, have a higher frequency response, and negative emotions, disgust and sadness, have a lower frequency response. Some of the spatial activation patterns resemble microstates C and D\cite{michel2018eeg} and new activation patterns not presented in microstate analysis were also learned (Fig.~\ref{fig3}). The building blocks of spatial transition patterns in EEG can be further investigated in future studies.

By employing contrastive learning for inter-subject alignment, the model successfully learns representations shared across subjects, ensuring robust generalization to new subjects. Other self-supervised learning techniques, like reconstruction-based pertaining\cite{wang2024dmmr}, have also been proven effective in mitigating cross-subject discrepancy. Future studies can further explore these techniques comprehensively.

\section{Conclusion}
In conclusion, we proposed a dynamic attention mechanism for EEG state transition modeling. The model achieved state-of-the-art cross-subject emotion recognition performance on three publicly available datasets. It obtained an accuracy of $75.4\pm5.5\%$ on binary classification and $59.3\pm7.7\%$ on nine-class classification on the FACED dataset, $88.1\pm3.6\%$ on the SEED dataset, and $73.6\pm12.7\%$ on the SEED-V dataset. The dynamic attention design outperforms strong competitive architectures including global channel attention and Transformer layers. The interpretation of the model can provide valuable insights into neural dynamics underlying emotion processing and the relationship between discrete emotions.


%

\ifCLASSOPTIONcompsoc
  \section*{Acknowledgments}
\else
  \section*{Acknowledgment}
\fi

This work was supported by the National Natural Science Foundation of China (U2336214, T2341003), Beijing Natural Science Foundation (IS23081), Shenzhen Science and Technology Innovation Committee (JCYJ20220818100213029, RCBS20231211090748082, KJZD20230923115221044), and IDG/McGovern Institute for Brain Research at Tsinghua University.

\ifCLASSOPTIONcaptionsoff
   \newpage
\fi

\bibliographystyle{IEEEtran}
\bibliography{Mybib}

%

\begin{IEEEbiography}[{\includegraphics[width=1in,height=1.25in,clip,keepaspectratio]{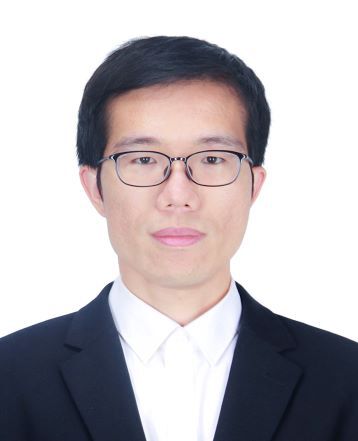}}]{Xinke Shen}
received his B.E. degree in
biomedical engineering from Beihang
University, Beijing, China, in 2017 and the
Ph.D. degree in biomedical engineering
at Tsinghua University, Beijing, China,
in 2023. He is currently a postdoctoral
fellow in the Department of Biomedical Engineering at Southern University of Science and Technology, Shenzhen, China. His research interests include affective computing and affective neuroscience.
\end{IEEEbiography}

\begin{IEEEbiography}[{\includegraphics[width=1in,height=1.25in,clip,keepaspectratio]{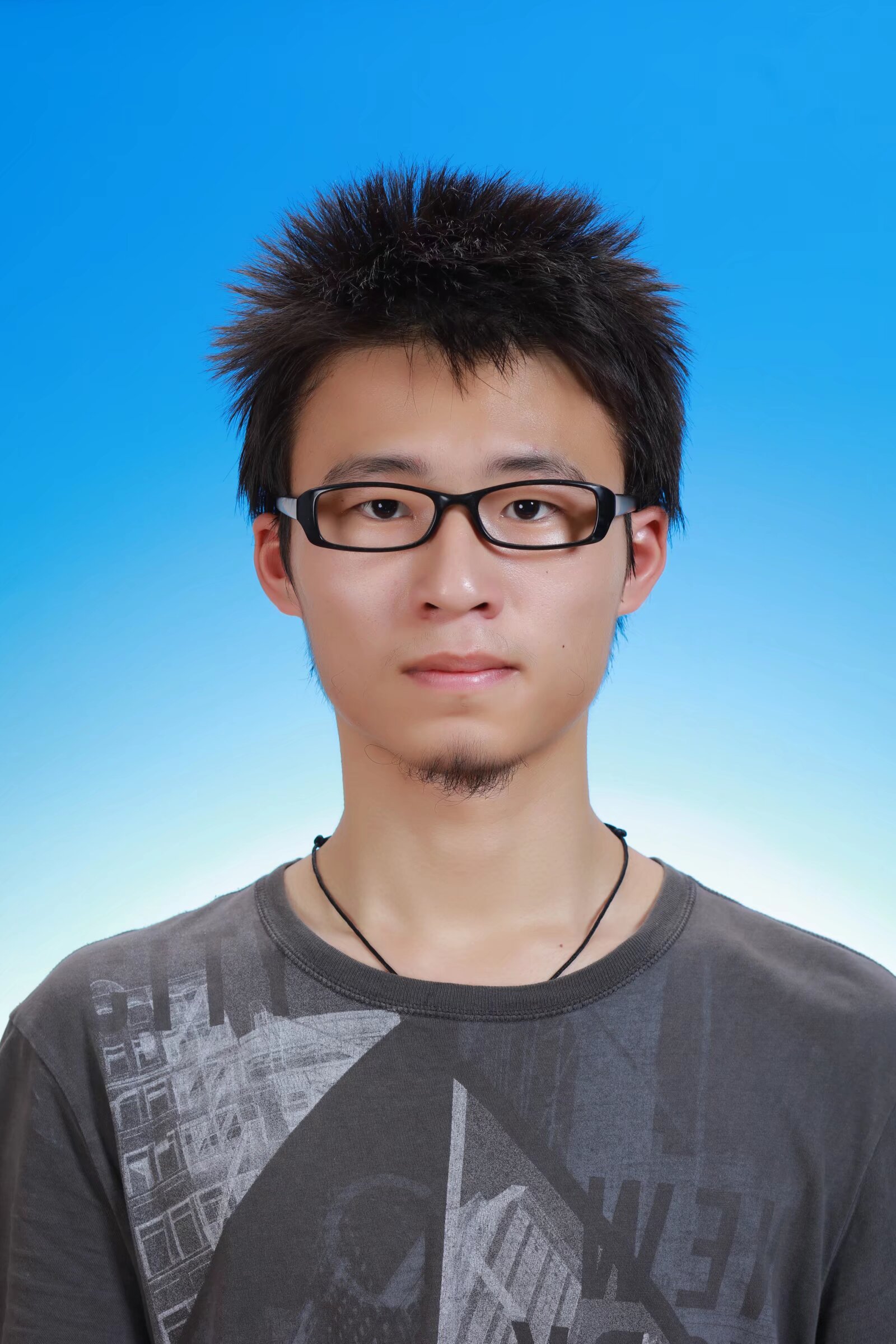}}]{Runmin Gan}
received his B.E. degree in electronic information engineering from Southwest Jiaotong University, Sichuan, China, in 2022. He is now working on the Ph.D. degree in biomedical engineering at Tsinghua University, Beijing, China. His research interests include computational neuroscience, affective computing, and the development of brain-inspired deep learning frameworks.
\end{IEEEbiography}

\begin{IEEEbiography}[{\includegraphics[width=1in,height=1.25in,clip,keepaspectratio]{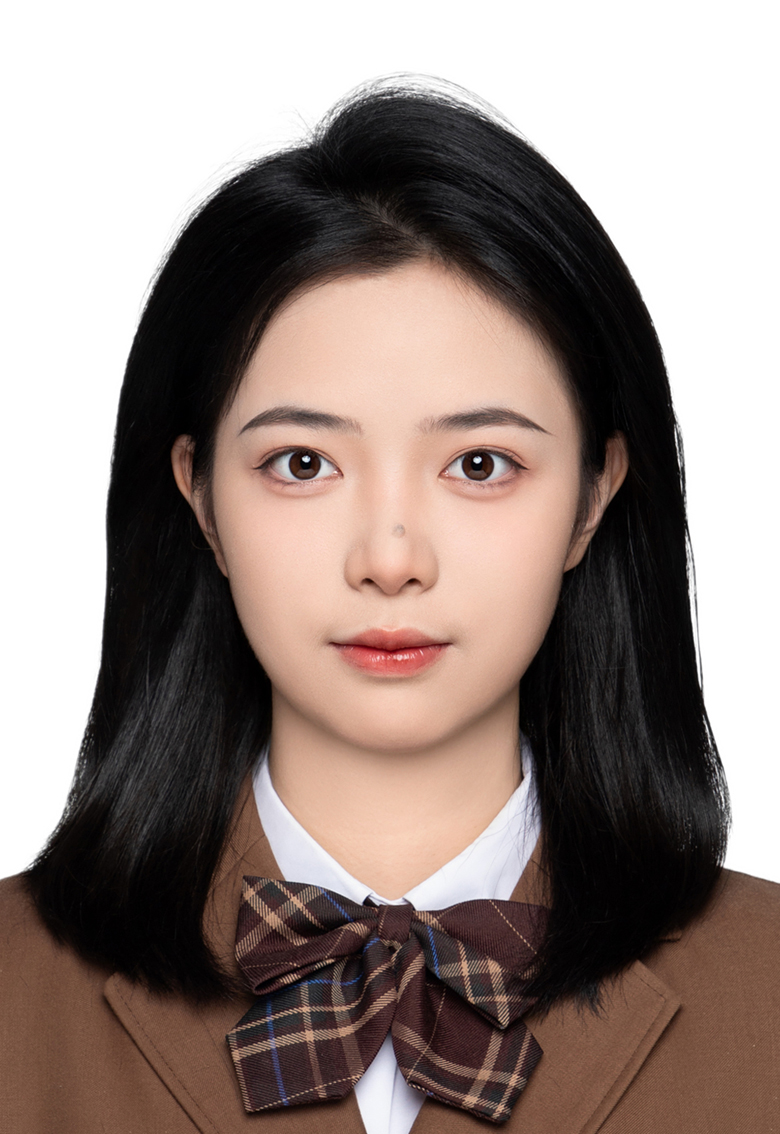}}]{Kaixuan Wang}
 is an undergraduate student in the School of Artificial Intelligence at Beijing Normal University, China. Her research interests include neuroinformatics, affective neuroscience, and intelligent systems for psychotherapy.
\end{IEEEbiography}

\begin{IEEEbiography}[{\includegraphics[width=1in,height=1.25in,clip,keepaspectratio]{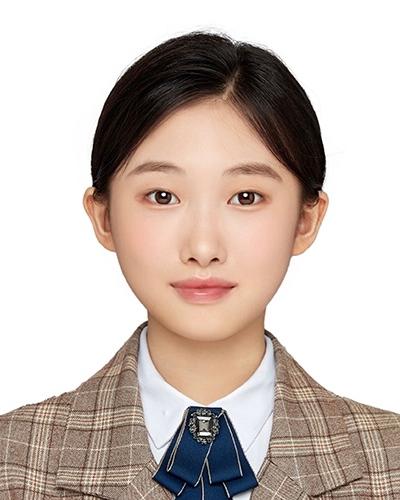}}]{Shuyi Yang}
 is an undergraduate student in the Department of Biomedical Engineering, Tsinghua University, Beijing, China. Her research interests include deep learning and the interpretability of clinical prediction models.
\end{IEEEbiography}

\begin{IEEEbiography}[{\includegraphics[width=1in,height=1.25in,clip,keepaspectratio]{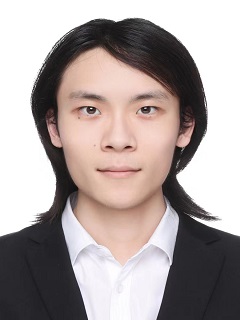}}]{Qingzhu Zhang}
is a master's student in the Department of Biomedical Engineering at Southern University of Science and Technology, Shenzhen, China. His research interests include affective computing and brain-computer interfaces.
\end{IEEEbiography}

\begin{IEEEbiography}[{\includegraphics[width=1in,height=1.25in,clip,keepaspectratio]{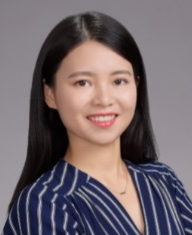}}]{Quanying Liu}
received her B.S. degree in electrical engineering and her M.S. degree in computer science at Lanzhou University in 2010 and 2013. After, she received her Ph.D. degree from Eidgenössische Technische Hochschule (ETH) Zurich and received postdoctoral training at the California Institute of Technology (CalTech). She is an Assistant Professor at the Department of Biomedical Engineering, Southern University of Science and Technology, and the principal investigator (PI) of the Neural Computing and Control Laboratory (NCC lab). Her research interests include multi-modal neural signal processing, AI for neuroscience, and optimization for neuromodulation.
\end{IEEEbiography}

\begin{IEEEbiography}[{\includegraphics[width=1in,height=1.25in,clip,keepaspectratio]{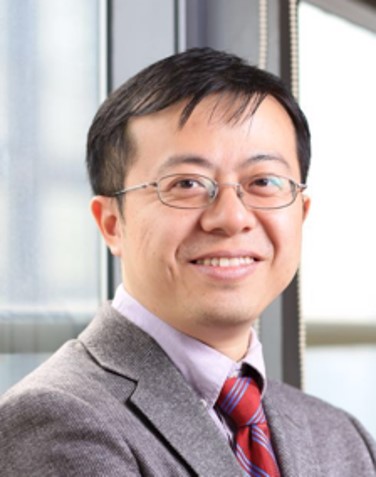}}]{Dan Zhang}
received his B.E. degree in automation 
in 2005 and his Ph.D. degree in biomedical engineering in 2011, both from Tsinghua University, Beijing, China. He was a postdoctoral fellow at the School of Medicine, Tsinghua University from 2011 to 2013. He is currently an associate professor at the Department of Psychological and Cognitive Sciences, Tsinghua University, Beijing, China. His research interests include social neuroscience, engineering psychology, and brain-computer interfaces. He is a member of the IEEE.
\end{IEEEbiography}

\begin{IEEEbiography}[{\includegraphics[width=1in,height=1.25in,clip,keepaspectratio]{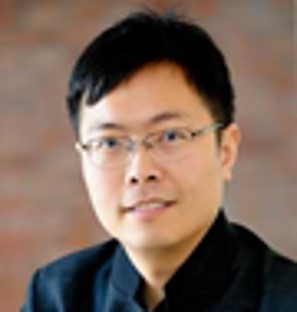}}]{Sen Song}
received his B.A. degree from University 
of Mississippi, US, in 1996 and his Ph.D. degree from Brandeis University, US, in 2002. He was a postdoctoral fellow at Cold Spring Harbor Laboratory, US, from 2002 to 2004 and at Massachusetts Institute of Technology, US, from 2004 to 2010. He 
is currently a principal investigator at the School of Biomedical Engineering at Tsinghua University. His research 
interests include brain-inspired neural networks, computational neuroscience, and affective computing.
\end{IEEEbiography}





\end{document}